\documentclass[twocolumn,aps,showpacs]{revtex4} %\documentstyle[twocolumn,aps]{revtex}

\begin{document} \title{Protein induced morphological transitions in KCl
crystal growth }

\author{B\'alint Szab\'o and Tam\'as Vicsek}

\affiliation{Department of Biological Physics, E\"otv\"os University, Budapest, P\'azm\'any P\'eter s\'et\'any
1/A, 1117 Hungary }

\date{\today}

\begin{abstract} We investigated the formation of KCl crystals on glass surface
by phase contrast, fluorescent, and atomic force microscopy on the micrometer scale, and observed interesting
morphological transitions as a function of the experimental conditions. The presence of proteins in the solution
from which the salt crystals grow during the drying up leads to complex microscopic patterns of crystals some of
which are analogous to those commonly observed on the macroscopic scale. We tested the effect of tubulin,
FITC-labeled albumin and IgG on the morphology of crystals grown either slowly or fast. A rich variety of protein
specific and concentration dependent morphologies was found and described by a morphological diagram. We give a
phenomenological interpretation, which can explain the growth of complex patterns. Fluorescent images prove that a
protein layer covers the surface of the KCl structures. We propose that this layer reduces the anisotropy of the
effective surface tension during growth. The tip splitting fractal regime is attributed to the decrease of
anisotropy. Other possible mechanisms, which can cause morphological transition, are also discussed. We found
elongated saw-toothed crystals induced by proteins, especially IgG and identified their structure. \end{abstract}

 \maketitle

\section{Introduction } The effect of proteins, especially albumin and IgG on crystal growth has biological
relevance. In crystal induced arthritis e.g., gout proteins bind the crystals \cite{a1} and have significant
impact on crystal growth \cite{a2}. The immune response to the appearance of crystals is driven by the specific
IgG-crystal interaction. This interaction stimulates crystal formation \cite{a3,a4,a5}. Although the rich
morphology of several inorganic macroscopic crystals grown on the surface of gels (i.e., in the presence of
protein) has been studied and discussed \cite{a6,a7} up to our knowledge this is the first study of pattern
formation induced by crystal-protein interactions. Moving unstable interfaces typically lead to the formation of
complex patterns. Surface tension of the boundary between the growing and surrounding phases is essential in
pattern formation \cite{a8,a9,a10,a11}. A growth mode of crystals covered with a thin layer has been observed and
explained by the surface tension between the different phases in a metal system \cite{a12}. The growth of
nonequilibrium interfaces covered by a thin surfactant film was both experimentally and theoretically studied
\cite{a13}. The anisotropy of the surface tension has crucial role. Viscous fingering, crystallization,
electrochemical deposition, and some other phenomena can be relatively well described by Laplacian growth with
appropriate boundary conditions. We argue that the growth of the numerous different microscopic patterns we
observed in our KCl crystallization experiments in the presence of proteins can be explained by the change of the
boundary conditions of the moving interface: the anisotropy of its surface tension is decreased by proteins.

\section{Materials and methods } 5 $\mu$l drops of 5 mg/ml KCl solution containing either tubulin (prepared
from bovine brain \cite{a14}) or FITC-BSA (bovine serum albumin conjugated with FITC, SIGMA) or human IgG-FITC
(immunoglobulin G conjugated with FITC, SIGMA) or no protein were placed on clean glass cover slips. We prepared
samples either by slow drying at $\sim$40 $^{\circ}$C in air or by fast drying in a strong airflow. Slow drying
took $\sim$5 minutes, airflow dried the surface on the scale of seconds. Bright field, phase contrast and
fluorescent images were acquired by an inverse Leica DM IRB microscope, 40x objective and a Nikon Coolprix 700
digital camera mounted on the microscope by homemade optical coupling. Atomic force microscopy (AFM) images of
tubulin coated crystals were captured with a commercial AFM (TopoMetrix Explorer, Santa Clara, CA) in contact mode
with a soft silicon nitride cantilever (Thermomicroscopes, coated sharp microlevers, model No. MSCT-AUHW, with
typical force constant 0.03 N/m, 20 nm nominal radius of curvature) under ambient conditions.

\section{Results}

\subsection{Tubulin } The fast drying of 0.5 mg/ml tubulin containing drops of 5 mg/ml KCl solution on glass
surface with air-flow results in a heterogeneous population of patterns that we investigated by AFM and phase
contrast microscopy. The morphology of KCl crystals can be similar to diffusion-limited aggregates known to have
fractal geometry. Dendritic patterns with stable and unstable tips are typical. FIG.\ \ref{fig1} shows a pattern,
which began to grow with stable tips, 4-fold symmetry. After a while tip splitting occurs at all the 4 tips.
Anisotropy drops dramatically at this point. The slow drying of drops containing 0.05 mg/ml tubulin gives
elongated saw-toothed crystals similar to the ones induced by IgG discussed below. In high concentration tubulin
inhibits crystal growth, an apparently homogeneous stain can be observed after the drop dries up.

\subsection{FITC-BSA } We tried to shed light on the role of protein in pattern formation of salt crystals with
the use of FITC labeled proteins. In case of the fast drying procedure 5 mg/ml FITC-BSA had similar effect to that
of tubulin. Various branching morphologies were found. Tip splitting caused fractal growth (FIG.\ \ref{fig2}). The
comparison of bright field and phase contrast images with the fluorescent ones show that crystals are covered by a
layer of albumin. Protein concentration dependence of the morphology was studied with the slow drying technique.
Under a critical protein concentration ($\sim$1$\mu$g/ml) only blocks of rectangular prism-shaped crystals grew
corresponding to the protein free case. In the concentration range of 1-1000 $\mu$g/ml cubic crystal centered
structures formed (FIG.\ \ref{fig3}). This observation reinforce that during a single growth process the symmetry
(anisotropy) of the pattern can dramatically change due to the change of the local conditions. As the growth
process elongates, the initial 4-fold symmetry with high anisotropy disappears and rather isotropic growth takes
place. In the higher protein concentration range dendritic crystals grew with stable tips (FIG.\ \ref{fig4}).

\subsection{IgG-FITC } To study the protein specificity of the phenomena we examined the impact of human IgG-FITC
on the KCl crystal growth. After the fast drying procedure using 2 mg/ml IgG the typical pattern we found was the
elongated saw-toothed crystal (FIG.\ \ref{fig5}). Branching structures can be observed. The angle between the
branches is $\sim$70$^{\circ}$. Concentration dependence of the morphology was investigated by the slow drying
method. Under $\sim$20 $\mu$g/ml only blocks of the prism-shaped crystals grew. Above this concentration up to
$\sim$1 mg/ml cubic and saw-toothed crystal centered structures were observed with an isotropic surrounding
pattern. FIG.\ \ref{fig6} shows the center of a typical one. In the higher concentration range dendritic patterns
formed with stable tips.

\subsection{Equilibrium shape} The equilibrium shape of KCl crystals after the slow
drying method using 10 $\mu$l 5 mg/ml KCl with 5 mg/ml FITC-BSA was observed by dropping 5 $\mu$l saturated KCl
solution on the crystals and covering the droplet with a cover slip. We found rounded forms instead of the normal
rectangular shapes of KCl. (Image not shown.)

\subsection{Gold substrate} We also studied the effect of the
substrate surface, which was glass in the above experiments. However, IgG gave similar results on gold substrate
to those on glass in case of the slow drying method, albumin induced the appearance of amorphous protein
aggregates and blocks of prism-shaped crystals on gold instead of the patterns described above. This fact
indicates the protein and substrate specificity of the phenomenon.

\section{Discussion } Proteins have significant
impact on KCl crystallization. The presence of proteins in the solution from which KCl crystals grow during the
drying up leads to the formation of protein specific and concentration dependent complex patterns, which we
described by a morphological diagram (FIG.\ \ref{fig7}). We can give a partial explanation of the diagram.
Meanwhile, the solution dries up the protein precipitates due to the increasing ionic strength. This process is
known as salting out. A thick layer of protein aggregate (precipitate) covers the surface of crystals. This cover
means 2 interfaces: one between the KCl crystal and the protein layer and one between the protein layer and the
liquid. While the surface tension of the former one $\gamma$$_{c-p}$ is anisotropic, that of the latter one
$\gamma$$_{p-l}$ is considered to be isotropic. The total surface tension $\gamma$$_t$ of the interface between
the growing and surrounding phases is given by the sum of these two terms:

\begin{equation}
\gamma_{t}=\gamma_{c-p} + \gamma_{p-l} \end{equation}

The value of $\gamma$$_{p-l}$ i.e., the isotropic term is significant since precipitation means insolubility of
the protein. The rounded equilibrium shape of crystals with protein cover reinforces the lower anisotropy of
surface tension. In the absence of protein the extra isotropic term corresponding to the protein-liquid interface
is missing. If \begin{equation} \gamma_{c-p} < \gamma_{p-l},
\end{equation} then an asymmetry arises between the two faces of the protein layer, which tends to bend it towards
the liquid phase favoring dendritic growth. We propose that tip splitting and fractal growth instead of the
formation of single crystals with 4-fold symmetry and stable tips can be attributed to the reduced anisotropy of
the surface tension in case of the fast growth. Similarly, slowly grown single crystal centered structures are
explained by the decrease of anisotropy of surface tension during the growth process. Our results demonstrate that
the level of anisotropy can change dramatically during the growth of a single pattern. This is likely to be caused
by the increase and the reaching of a critical level of protein precipitation as the drop is drying up. Various
examples of pattern formation driven by moving unstable interfaces can be described by Laplacian growth. In this
case the boundary condition along the $\Gamma$ interface containing the dimensionless u$_\Gamma$ concentration
term is \cite{a8,a9,a10,a11}:

\begin{equation} u_{\Gamma} = \Delta -d_{0}(\Theta)\kappa -\beta(\Theta)v_{n},
\end{equation}

where $\Delta$ is the undercooling, the capillary length d$_0$ is proportional to the surface tension, $\kappa$
denotes the local curvature of the interface, $\beta$ is the kinetic coefficient, $\Theta$ is the angle between
the normal to the surface and a fixed crystallographic direction, v$_{n}$ is the normal velocity of the interface.
To explain the stable tip-tip splitting transition, we propose that d$_0$ is altered and its anisotropy is
decreased due to the adherence of the protein layer to the KCl crystal surface. Other effects may also have an
impact on the morphology of KCl patterns. The lower diffusion coefficient of K$^+$ and Cl$^-$ ions in the liquid
phase due to the high protein concentration during drying up might induce transport limited growth. Both the
decreased velocity of the interface and the modification of the angular dependence of its coefficient ($\beta$) by
a rather isotropic incorporation process of KCl into the crystal caused by the protein coating can lead to lower
anisotropy of the pattern. The incorporation of proteins into calcite crystals is known to have an effect on the
morphology \cite{a15}. However, fluorescent images displayed high protein concentration on the surface of
crystals, we can not totally exclude incorporation. IgG typically and under some conditions tubulin and albumin
also induced the growth of elongated saw-toothed crystals instead of rectangular needle-shaped ones. Branching
saw-toothed structures were also found, the angle between the branches is $\sim$70$^{\circ}$. We argue that these
are single crystals elongated in the (111) direction of the cubic crystal i.e., their axis lies in this direction.
The angle between (111) and directions with the same symmetry e.g. (11-1) is 70.5$^{\circ}$. The mechanism leading
to the formation of these crystals is unknown. Proteins can adhere to specific faces of the crystal with increased
affinity \cite{a2}. This anisotropic interaction influences the anisotropy of the surface tension at the
crystal-protein interface $\gamma$$_{c-p}$, which may result in the change of the direction of the fast growing
tip. The background of the protein specific behavior is unclear. As the experiment on the gold substrate
indicates, in addition to the protein-crystal interaction protein-substrate and crystal-substrate effects also
should be considered in further studies.

\section{Acknowledgments} This work was supported by the Hungarian
National Scientific Research Fund (OTKA, No. T-034995). We thank No\'emi Rozlosnik and Andr\'as Czir\'ok for the
help in establishing the instrumental background of our experiments, and Judit Ov\'adi and Emma Hlavanda for the
tubulin samples.

\begin{figure} \caption{80x80 $\mu$m$^{2}$ deflection mode AFM image of tubulin induced patterns
of KCl after the fast drying procedure. Note the transition of the initially 4-fold symmetric crystallization to
isotropic fractal growth at all the 4 tips of the structure caused by consecutive tip splittings.} \label{fig1}
\end{figure}

\begin{figure} \caption{135x135 $\mu$m$^{2}$ fluorescent image of 2 patterns of KCl crystals with
different fractal dimensions grown in the presence of FITC-BSA with the fast drying method. Brightness is
proportional to the concentration of FITC-BSA. The comparison with phase contrast and bright field images prove
that albumin covers the crystals.} \label{fig2} \end{figure}

\begin{figure} \caption{135x135 $\mu$m$^{2}$
fluorescent image of a cubic crystal centered structure of KCl crystals induced by FITC-BSA using the slow drying
method. The anisotropic single crystal in the center is surrounded by an isotropic pattern.} \label{fig3}
\end{figure}

\begin{figure} \caption{135x135 $\mu$m$^{2}$ fluorescent image of stable tipped dendritic KCl
crystals induced by high concentration of albumin.} \label{fig4} \end{figure}

\begin{figure} \caption{135x135
$\mu$m$^{2}$ fluorescent image of KCl crystals with saw-toothed shape grown with IgG by the fast drying procedure.
The angle between the branches is $\sim$70$^{\circ}$. The bright contour of the crystals corresponds to the
IgG-FITC cover on their surface.} \label{fig5} \end{figure}

\begin{figure} \caption{205x205 $\mu$m$^{2}$
fluorescent image. Saw-toothed crystals are shown in the center of a pattern, which is isotropic in the
surrounding regions. This sample was prepared with the slow drying procedure in the presence of IgG.}
\label{fig6}\end{figure}

\begin{figure} \caption{Morphological diagram of the protein induced patterns of KCl
crystals.} \label{fig7} \end{figure}

 \end{document}